\documentclass[aps,preprint,groupedaddress,epsf,tighten]{revtex4}

\usepackage{graphicx}
\usepackage{amsmath}
\usepackage{bm}

\newcommand{\vsl}{v\!\!\!\slash}

\newcommand{\Dsl}{D\!\!\!\!\slash\,}
\newcommand{\Dslp}{D\!\!\!\!\slash_\perp\,}

\begin{document}

 
{\tighten
\preprint{\vbox{\hbox{TECHNION 96-15}
                \hbox{MZ-TH/96-17} }}
 
\title{Spectator Effects in the Heavy Quark Effective Theory}
 
\author{B.~Blok, J.~G.~K\"orner, 
D.~Pirjol and J.~C.~Rojas\footnote{On leave
of absence from the Universidad de Santiago de Chile, Avenida Bernardo O'Higgins 3363,
Santiago, Chile}}

\affiliation{Department of Physics, Technion-Israel Institute of Technology, 
Haifa 32000, Israel}

\affiliation{Johannes-Gutenberg Universit\"at, Institut f\"ur Physik (ThEP), Staudinger Weg 7,
         55099 Mainz, Germany}

\begin{abstract}%
We present a complete analysis of the Heavy Quark Effective Theory Lagrangian at order $1/m^2$
in the leading logarithmic approximation, including effects induced by spectator quarks.
At this order new correction terms appear in the effective Lagrangian, as
four-quark operators containing both heavy and light quark fields. We compute the coefficients
of these operators in the leading-logarithmic approximation, and the matching conditions
at one-loop order. Two of these operators break the heavy
quark spin symmetry, and we estimate their contribution to the hyperfine splitting of the heavy
mesons. We find that they make a positive contribution to
the hyperfine splitting of about 10\% of the measured splitting in the charm case and of up
to 5\% in the bottom case.

\end{abstract}

\maketitle

}

\newpage

\section{Introduction}

The heavy mass expansion established itself as a valuable tool for the study of hadrons containing
one heavy quark \cite{IW,Ne,Sh}. This expansion is formulated most naturally in terms of an effective
theory, the
heavy quark effective theory (HQET), which is an approximation to QCD with one heavy quark.
The corrections to this approximation are controlled by the small parameter $\Lambda_{QCD}/m$, where
$\Lambda_{QCD}\simeq 300$ MeV is a typical scale of low-energy QCD and $m$ is the heavy quark mass.

The applications of the heavy quark effective theory to physical problems presented so far
include correction terms of order $1/m^2$ \cite{FN,Man,BSUV} and in some recent instances even
$1/m^3$ \cite{BDS,GK}. 
We investigate in this paper the effects induced by spectator quarks in the HQET.
They appear first at order $1/m^2$ and are associated
in an effective theory language
with 4--quark operators containing both heavy- and light-quark fields.
These operators mix under renormalization with the other operators of dimension six present
in the HQET Lagrangian already at tree level, and their inclusion changes the form of the 
renormalized Lagrangian.

We define the operator basis in Section II and compute their mixing under renormalization.
The coefficients of these operators can be obtained from a one-loop matching calculation, which
is presented in Section III, together with a leading-log renormalization-group running.
As an application we estimate in Section IV their total contribution to the
hyperfine splitting of heavy-light mesons in the framework of the factorization approximation.

\section{Dimension-6 Operators}

The Lagrangian of the heavy quark effective theory is written as an expansion in the inverse
heavy quark mass \cite{IW,Ne,Sh}
\begin{equation}\label{1}
{\cal L}_{HQET} = \bar h(iv\cdot D)h + \frac{1}{2m}{\cal L}_1 + \frac{1}{(2m)^2}{\cal L}_2 + \cdots
\end{equation}
The heavy quark field $h$ annihilates static heavy quarks moving with a fixed velocity $v$ and
satisfies
the condition $\frac12(1+\vsl)h = h$. The finite-mass effects appear as interaction terms in
the Lagrangian suppressed by powers of $1/m$. The first two correction terms have the well-known
expressions \cite{L,EH,FGL,KT,BKP}
\begin{eqnarray}\label{2}
{\cal L}_1 &=& c_k(\mu) \bar h(iD)^2h + c_m(\mu) \bar h\frac{g}{2}\sigma_{\mu\nu}F^{\mu\nu}h
+ c_e(\mu)\bar h(iv\cdot D)^2h\\
{\cal L}_2 &=& c_D(\mu) {\cal O}_D + c_{SO}(\mu) {\cal O}_{SO}\,.\label{3}
\end{eqnarray}
The coefficients of the operators in Eq.~(\ref{2}) are known in the leading
logarithmic approximation \cite{EH,FGL}
\begin{eqnarray}\label{6}
c_k(\mu) = 1\,,\qquad c_m(\mu) = -\left(\frac{\alpha_s(\mu)}{\alpha_s(m)}\right)^{-3/\beta_0}
\,,\qquad c_e(\mu) = 2-3\left(\frac{\alpha_s(\mu)}{\alpha_s(m)}\right)^{-8/(3\beta_0)}
\end{eqnarray}
where $\beta_0=11-\frac23n_f$ is the one-loop coefficient of the beta function for $n_f$ active
quarks. The coefficient of the last operator in (\ref{2}) is gauge-dependent; the value
quoted above corresponds to the Feynman gauge.

To order $1/m^2$ there are two local operators contributing at tree-level, the Darwin term and the
spin-orbit interaction energy respectively
\begin{eqnarray}\label{4}
{\cal O}_D &=& \frac{g}{2}(D^\mu F_{\mu\nu}^a)(\bar hv_\nu t^ah)\\\label{5}
{\cal O}_{SO} &=& ig(\bar h\sigma_{\alpha\nu}v_\mu F^{\mu\nu}D^\alpha h) +
\frac{ig}{2} (D_\alpha F^a_{\mu\nu})(\bar h\sigma^{\alpha\nu} v^\mu t^ah)\,.
\end{eqnarray}
The coefficients $c_D(\mu)$ and $c_{SO}(\mu)$ were calculated in \cite{BKP} (see also \cite{CKO,BO})
using a different operator basis and in the absence of light quarks. We reconsider here their
calculation in a basis more suited for our present purposes, with a different result for $c_D(\mu)$.

In addition to the operators shown in (\ref{3}), there are three other types of dimension-6
operators which must be added to ${\cal L}_2$:
\begin{enumerate}
\item[a)] the gluonic operators
\begin{equation}\label{7}
{\cal O}_{F^3} = \frac14 gf_{abc}F_{\mu\nu}^a F_{\nu\lambda}^b F_{\lambda\mu}^c\,,\qquad
{\cal O}_{(DF)^2} = \frac12 (D^\mu F^a_{\mu\nu} ) (D_\lambda F^{a\lambda\nu} )\,.
\end{equation}
A third gluonic operator $F^a_{\mu\nu} D^2 F^{a\mu\nu}$ has been eliminated
in zero momentum insertions with the help of the Bianchi identity, which gives
\begin{eqnarray}
\int \mbox{d}^4x (4{\cal O}_{F^3} + {\cal O}_{(DF)^2} + \frac14
F^a_{\mu\nu} D^2 F^{a\mu\nu}) = 0
\end{eqnarray}
Their coefficients have been computed many times~\cite{Shifman,BKP,CS}. 
At one-loop order they are given by
\begin{eqnarray}\label{gl0}
c_{F^3}(m) = \frac{\alpha_s(m)}{45\pi}\,, \qquad
c_{(DF)^2}(m) = -\frac{2\alpha_s(m)}{15\pi}
\end{eqnarray}

At one-loop order ${\cal O}_{F^3}$ does not mix with other dimension-6 operators and
will be neglected in the following. The second operator ${\cal O}_{(DF)^2}$ will be
eliminated in favor of 4-quark operators using the equation of motion of the gluon
field (see Eq.~(\ref{eqm1}) below).

\item[b)] Operators which vanish by the equation of motion of the heavy quark field $iv\cdot Dh=0$.
Even though their expectation values vanish, they can contribute when considering mass corrections 
to the matrix elements of currents. A discussion of their renormalization is given in Appendix
A.

\item[c)] 4-quark operators built out of heavy-light and light-light quark fields.
It is these operators which will be the main point of interest of this paper.
\end{enumerate}

There are four independent heavy-light 4-quark operators, which can be conveniently chosen as
follows
\begin{eqnarray}\label{8}
{\cal O}_1^{hl} &=& \frac{g^2}{2}\sum_q(\bar q\gamma_\mu t^a q)(\bar h\gamma^\mu t^a h)\\
\label{9}
{\cal O}_2^{hl} &=& \frac{g^2}{2}\sum_q(\bar q\gamma_\mu\gamma_5t^aq)(\bar h\gamma^\mu\gamma_5t^ah)\\
{\cal O}_3^{hl} &=& \frac{g^2}{2}\sum_q(\bar q\gamma_\mu q)(\bar h\gamma^\mu h)\\
{\cal O}_4^{hl} &=& \frac{g^2}{2}\sum_q(\bar q\gamma_\mu\gamma_5 q)(\bar h\gamma^\mu\gamma_5 h)\,.
\label{11}
\end{eqnarray}
The summation extends over $n_f$ dynamic quarks, for example
$n_f=4$ $(q=u,d,s,c)$ for the case of a heavy bottom quark.

The equation of motion for the gluon field gives a relation among the Darwin 
term ${\cal O}_D$ Eq.~(\ref{4}) and one of the 4-quark operators ${\cal O}_i^{hl}$
\begin{eqnarray}\label{eqm}
{\cal O}_D = {\cal O}_1^{hl} + \frac{g^2}{2} (\bar h t^a h)(\bar h t^a h)\,.
\end{eqnarray}

The structure of the possible 4-quark operators containing only light quark fields is 
similar to that of the heavy-light operators. There are altogether 10 such operators,
which can be divided into three groups with different transformation properties under
$SU(n_f)_L \times SU(n_f)_R$. The operators in each group mix only among themselves 
under renormalization. Their renormalization has been discussed in detail in
Refs.~\cite{SVZ,JK,BB}. Only one group of operators is relevant for our case, which
can be chosen as
\begin{eqnarray}\label{12}
{\cal O}_1^{ll} &=& \frac{g^2}{2}\sum_{q,q'}
(\bar q\gamma_\mu t^aq)(\bar q\gamma^\mu t^aq)\\
{\cal O}_2^{ll} &=& \frac{g^2}{2}\sum_{q,q'}
(\bar q\gamma_\mu\gamma_5t^aq)(\bar q\gamma^\mu\gamma_5t^aq)\\
{\cal O}_3^{ll} &=& \frac{g^2}{2}\sum_{q,q'}
(\bar q\gamma_\mu q)(\bar q\gamma^\mu q)\\
{\cal O}_4^{ll} &=& \frac{g^2}{2}\sum_{q,q'}
(\bar q\gamma_\mu\gamma_5 q)(\bar q\gamma^\mu\gamma_5 q)\,.
\label{15}
\end{eqnarray}

The complete basis of the dimension-6 operators includes also three nonlocal operators consisting
of time-ordered products of dimension-5 operators
\begin{eqnarray}\label{16}
{\cal O}_{kk} &=& \frac{i}{2}\int\mbox{d}^4x\, \mbox{T}[\bar h(iD)^2h](x)\,
[\bar h(iD)^2h](0)\\
{\cal O}_{km} &=& i\int\mbox{d}^4x\, \mbox{T} [\bar h(iD)^2h](x)\,
[\bar h\frac{g}{2}\sigma\!\cdot\! Fh](0)\\
{\cal O}_{mm} &=& \frac{i}{2}\int\mbox{d}^4x\, \mbox{T} [\bar h\frac{g}{2}\sigma\!\cdot\! Fh](x)
\,[\bar h\frac{g}{2}\sigma\!\cdot\! Fh](0)\,.
\label{18}
\end{eqnarray}

We will use a compact vector notation for the thirteen operators
(\ref{4},\ref{5},\ref{8}-\ref{18}), defined as follows
\begin{equation}\label{19}
\hat {\cal O} = \left( \begin{array}{l}
{\cal H}\\
{\cal O}^{hl}\\
{\cal O}^{ll}\end{array} \right)\,,\qquad{\rm with}\,\,
{\cal H} = \left(\begin{array}{l}
{\cal O}_D\\
{\cal O}_{SO}\\
{\cal O}_{kk}\\
{\cal O}_{km}\\
{\cal O}_{mm}\end{array}\right)\,,\,\,\,
\hat {\cal O}^{hl} = \left(\begin{array}{l}
{\cal O}_1^{hl}\\
{\cal O}_2^{hl}\\
{\cal O}_3^{hl}\\
{\cal O}_4^{hl}\end{array}\right)\,,\,\,\,
\hat {\cal O}^{ll} = \left(\begin{array}{c}
{\cal O}_1^{ll}\\
{\cal O}_2^{ll}\\
{\cal O}_3^{ll}\\
{\cal O}_4^{ll}\end{array}\right)\,\,\,\,.
\end{equation}
and a corresponding vector notation for their Wilson coefficients $c_i(\mu)$, carrying
the same index as the operators ${\cal O}_i$.
Requiring that the total
renormalized Lagrangian ${\cal L}_2$ be scale-independent gives a renormalization-group equation
for the coefficients $\hat c(\mu)$
\begin{equation}\label{20}
\mu\frac{\mbox{d}}{\mbox{d}\mu}\hat c - \hat \gamma^T\hat c = 0\,.
\end{equation}
The anomalous dimension matrix can be written as
\begin{eqnarray}\label{21}
\hat\gamma = \left( \begin{array}{ccc}
A & B & 0 \\
C & D & 0 \\
0 & E & F \end{array}\right)\,.
\end{eqnarray}
To one-loop order the blocks in this matrix take the values
\begin{mathletters}
\label{22}
\begin{equation}\label{22a}
A = \frac{g^2}{(4\pi)^2}\left( \begin{array}{ccccc}
4 & 0 & 0 & 0 & 0 \\
0 & 0 & 0 & 0 & 0 \\
-\frac{146}{9} & 0 & 0 & 0 & 0 \\
0 & -12 & 0 & 6 & 0 \\
-10 & 0 & 0 & 0 & 12 \end{array}\right)\,,
\end{equation}
\begin{equation}\label{22b}
B = \frac{g^2}{(4\pi)^2}\left( \begin{array}{cccc}
9 & 0 & 0 & 0 \\
0 & -5 & 0 & -\frac83 \\
-18 & 0 & 0 & 0 \\
0 & -\frac{40}{3} & 0 & -\frac{64}{9} \\
9 & -\frac53 & 0 & -\frac89 \end{array}\right)\,,
\end{equation}
\begin{equation}\label{22c}
C = \frac{g^2}{(4\pi)^2}\left( \begin{array}{ccccc}
\frac43 n_f & 0 & 0 & 0 & 0 \\
0 & 0 & 0 & 0 & 0 \\
0 & 0 & 0 & 0 & 0 \\
0 & 0 & 0 & 0 & 0 \end{array}\right)\,,
\end{equation}
\begin{equation}\label{22d}
D = \frac{g^2}{(4\pi)^2}\mbox{diag}(13-\frac43 n_f, 13-\frac43 n_f,
22 - \frac43 n_f, 22-\frac43 n_f)\,,
\end{equation}
\begin{equation}\label{22e}
E = \frac{g^2}{(4\pi)^2}\left( \begin{array}{cccc}
-\frac49 + \frac83 n_f & 0 & 0 & 0 \\
-\frac49 & 0 & 0 & 0 \\
\frac83 & 0 & 0 & 0 \\
\frac83 & 0 & 0 & 0  \end{array}\right)\,,
\end{equation}
\begin{equation}\label{22f}
F = \frac{g^2}{(4\pi)^2}\left( \begin{array}{cccc}
\frac{113}{9} + \frac43 n_f & 5 & 0 & \frac83 \\
\frac{41}{9} & 13-\frac43 n_f & \frac83 & 0 \\
\frac83 & 12 & 22-\frac43 n_f & 0 \\
\frac{44}{3} & 0 & 0 & 22-\frac43 n_f \end{array}\right)\,.
\end{equation}
\end{mathletters}
The blocks $D$ and $F$ contain contributions which take into account the running of the $g$ factors
contained in the definition of the four-quark operators.
We have implicitly used the fact established in \cite{BKP} that the form of ${\cal O}_{SO}$ as
given in (\ref{5}) is preserved under renormalization. We agree with \cite{BO} on the renormalization
of the Darwin term in $A$ in the absence of the mixing with the four-quark operators.
The lower diagonal block $F$ has been calculated previously in \cite{CS}. ({\em Note added.} After
correcting the entries $C_{11},D_{11},E_{11}$ we agree with the results of 
Ref.~\cite{BM}. These changes affect only the result for the Darwin term $c_D(\mu)$,
but leave all the other results of the paper unchanged.)

\section{Matching and running}

The coefficients of the operators ${\cal H}$ in Eq.~(\ref{19}) are given, at the matching scale
$\mu=m$, by their tree-level values $c_{{\cal H}}(m) =$ column$(1,1,1,-1,1)$.
The coefficients of the non-local operators (\ref{16}-\ref{18}) are given simply by the products
of the respective local dimension-5 operators
\begin{eqnarray}\label{ckk}
c_{kk}(\mu) &=& 1\\
c_{km}(\mu) &=& -\left(\frac{\alpha_s(\mu)}{\alpha_s(m)}\right)^{-3/\beta_0}\\
c_{mm}(\mu) &=& \left(\frac{\alpha_s(\mu)}{\alpha_s(m)}\right)^{-6/\beta_0}\,.\label{cmm}
\end{eqnarray}

Integration of the RGE for the Wilson coefficient of the Darwin operator $c_D(\mu)$ is 
complicated by the mixing between ${\cal O}_D$ and ${\cal O}_1^{hl}$. 
The initial conditions for the leading log running of the Wilson coefficients of the 4-quark 
operators are $c_i^{hl}(m)=0$. We compute below also the leading nonvanishing contribution to the
matching of these Wilson coefficients, of order $\alpha_s(m)$ (see Eqs.~(\ref{26})-(\ref{29})). 
 
We start by solving the RGE for the coefficients $\hat C(\mu) = (c_D(\mu),c_1^{hl}(\mu))$,
which has the form
\begin{eqnarray}
\mu \frac{d}{d\mu} \hat C(\mu) = A_2^T \hat C(\mu) + \hat B(\mu)
\end{eqnarray}
with $A_2^T$ the corresponding $2\times 2$ block of the anomalous dimension matrix $\hat \gamma$ 
in Eq.~(\ref{21}), and $\hat B(\mu)$ an inhomogeneous term
\begin{eqnarray}
A_2^T = \frac{\alpha_s}{4\pi}\left(
\begin{array}{cc}
4 & \frac43 n_f \\
9 & 13 - \frac43 n_f \\
\end{array}
\right)\,,\qquad
\hat B(\mu) = 
\frac{\alpha_s}{4\pi}\left(
\begin{array}{c}
-\frac{146}{9} - 10 z^{-6/\beta_0} \\
- 18 + 9 z^{-6/\beta_0} \\
\end{array}
\right)
\end{eqnarray}
We denoted here $z = \alpha_s(\mu)/\alpha_s(m)$, and we neglected contributions
to $\hat B(\mu)$ from the light-light 4-quark operators ${\cal O}_i^{ll}$. These
contributions are of the form $\alpha_s(\alpha_s \log(\mu/m))^n$, and are thus
formally of higher order than the leading logs computed here. They are however
calculable, see below.

The matrix $A_2^T$ is diagonalized as
\begin{eqnarray}
U^{-1} A_2^T U = \frac{\alpha_s}{4\pi} \left(
\begin{array}{cc}
\gamma_1 & 0 \\
0 & \gamma_2 \\
\end{array}
\right)\,, \qquad \mbox{with} \quad U =
\left(
\begin{array}{cc}
\frac{4}{27}n_f & -1 \\
1 & 1 \\
\end{array}
\right)\,,\quad
U^{-1} = \frac{1}{R}
\left(
\begin{array}{cc}
1 & 1 \\
-1 & \frac{4}{27}n_f \\
\end{array}
\right)
\end{eqnarray}
with $\gamma_1 = 13, \gamma_2 = 4-\frac43 n_f$ and $R = 1 + \frac{4}{27}n_f$. 
Redefining the Wilson coefficients  as
 $\hat C(\mu) = U \hat C'(\mu)$,
the RGEs for the components of $\hat C'(\mu)$ are decoupled 
\begin{eqnarray}
\mu \frac{d}{d\mu} \hat C'(\mu) = U^{-1} A_2^T U \hat C'(\mu) + U^{-1} \hat B(\mu)
\end{eqnarray}
and can be solved by simple integration.

Solving the RGE for the Wilson coefficients $c_D(\mu), c_1^{hl}(\mu)$, one has to
distinguish the two cases i) $n_f \neq 3$, and ii) $n_f=3$ separately, corresponding
to $\gamma_2$ being nonvanishing, and vanishing, respectively.

The solutions for the case i) $n_f \neq 3$ are given by
\begin{eqnarray}\label{23a}
c_D(\mu) &=& \frac{1}{R} \left\{ \frac{16 n_f^2 + 96 n_f - 81}{108(n_f+6)} z^{-6/\beta_0}
- \frac{1232}{3159} n_f z^{-\gamma_1/(2\beta_0)} \right. \\
& &\qquad
+ \left. \frac{12 n_f^2 + 353 n_f + 1605}{12(n_f-3)(n_f+6)} z^{-\gamma_2/(2\beta_0)}
 +  \frac{77(16 n_f^2 - 48 n_f - 1053}{3159(n_f-3)} \right\}\nonumber \\
\label{23b}
c_1^{hl}(\mu)  &=& \frac{1}{R} \left\{ \frac{4 n_f + 27}{4(n_f+6)} z^{-6/\beta_0}
+ \frac{308}{117} z^{-\gamma_1/(2\beta_0)} \right. \\
& &\qquad
- \left. \frac{12 n_f^2 + 353 n_f + 1605}{12(n_f-3)(n_f+6)} z^{-\gamma_2/(2\beta_0)}
 -  \frac{77(4 n_f - 51}{117(n_f-3)} \right\}\nonumber 
\end{eqnarray}

The corresponding results for case ii) $n_f = 3$ are 
\begin{eqnarray}\label{24a}
c_D(\mu) &=& \frac{1}{R} \left\{ 
\frac{13}{36} z^{-6/\beta_0} 
- \frac{1232}{1053} z^{-13/(2\beta_0)} 
+  \frac{154}{9\beta_0} \log z + \frac{9491}{4212} \right\} \\
\label{24b}
c_1^{hl}(\mu)  &=& \frac{1}{R} \left\{
\frac{13}{12} z^{-6/\beta_0} 
- \frac{308}{117} z^{-13/(2\beta_0)} 
-  \frac{154}{9\beta_0} \log z + \frac{725}{468} \right\}
\end{eqnarray}

For both cases i) and ii), the sum of the two Wilson coefficients
is given by a common expression
\begin{eqnarray}\label{cD}
c_D(\mu) + c_{1}^{hl}(\mu) = z^{-6/\beta_0} - \frac{308}{117} z^{-13/(2\beta_0)} 
+ \frac{308}{117}\,.
\end{eqnarray}
This combination corresponds to the coefficient of the Darwin operator
in Ref.~\cite{BM}, in a basis in which the 4-quark operator ${\cal O}_1^{hl}$
is eliminated using the equation of motion Eq.~(\ref{eqm}). The result Eq.~(\ref{cD})
agrees with Eq.~(19) of Ref.~\cite{BM}\footnote{It also agrees with the sum 
$c_D+c_1^{hl}$ of the Eqs.~(27),(28) in the v1 of the present paper.}.

The results for $c_D(\mu), c_1^{hl}(\mu)$ presented above neglect mixing into
${\cal O}_1^{hl}$ from the light-light 4-quark operators ${\cal O}_i^{ll}$.
As mentioned above, they are formally next-to-leading order, and are comparable
with other contributions which can not be yet fully computed. They require
the one-loop matching of the Darwin operator (obtained in Ref.~\cite{Aneesh}), and the
2-loop anomalous dimension $A_2^T$ (not yet available).
However, the mixing of $c_i^{ll}$
into $c_D(\mu), c_1^{hl}(\mu)$ is a well defined effects and can be computed 
from our results.

We present next the Wilson coefficients $c_i^{ll}$ in leading log approximation.
The equation of motion for the gluon field relates the operator ${\cal O}_{(DF)^2}$
to the 4-quark operators
\begin{eqnarray}\label{eqm1}
{\cal O}_{(DF)^2} = \frac12 (D^\mu F^a_{\mu\nu} ) (D_\lambda F^{a\lambda\nu} )
= 2 {\cal O}_1^{hl} + {\cal O}_1^{ll} + \frac{g^2}{2} (\bar h t^a h)(\bar h t^a h)\,.
\end{eqnarray}
We will use this relation to eliminate the gluonic operator in favor of the
4-quark operators ${\cal O}_1^{hl}$ and ${\cal O}_1^{ll}$. Using Eq.~(\ref{gl0}), 
this gives the initial condition for $c_1^{ll}(m)$ 
\begin{eqnarray}
c_1^{ll}(m) = - \frac{2\alpha_s(m)}{15\pi}\,.
\end{eqnarray}
It is now straightforward to solve the RGE for the coefficients $c_i^{ll}(\mu)$.
We focus on the physically interesting case of the $1/m_c^2$ corrections, and take
as thus $n_f=3$. The results are given by (with $\beta_0 = 9$)
\begin{eqnarray}
c_1^{ll}(\mu) &=& (0.0246 z^{-\delta_1/(2\beta_0)} + 0.4359 z^{-\delta_2/(2\beta_0)}
+ 0.1646 z^{-\delta_3/(2\beta_0)} + 0.3748 z^{-\delta_4/(2\beta_0)}) c_1^{ll}(m) 
\nonumber\\
c_2^{ll}(\mu) &=& (-0.0434 z^{-\delta_1/(2\beta_0)} + 0.1961 z^{-\delta_2/(2\beta_0)}
- 0.3764 z^{-\delta_3/(2\beta_0)} + 0.2237 z^{-\delta_4/(2\beta_0)}) c_1^{ll}(m) 
\nonumber\\
c_3^{ll}(\mu) &=& (-0.051 z^{-\delta_1/(2\beta_0)} + 0.0767 z^{-\delta_2/(2\beta_0)}
+ 0.0741 z^{-\delta_3/(2\beta_0)} - 0.0998 z^{-\delta_4/(2\beta_0)}) c_1^{ll}(m) 
\nonumber\\
c_4^{ll}(\mu) &=& (0.0290 z^{-\delta_1/(2\beta_0)} + 0.1706 z^{-\delta_2/(2\beta_0)}
- 0.0324 z^{-\delta_3/(2\beta_0)} - 0.1672 z^{-\delta_4/(2\beta_0)}) c_1^{ll}(m) 
\nonumber\\
\end{eqnarray}
The eigenvalues of the mixing matrix $F$ are given by $\alpha_s/(4\pi) \delta_i$,
with $\delta_1 = 20.2680, \delta_2 = 24.8138, \delta_3 = 4.4516, \delta_4 = 12.0222$.

We turn next to the spin symmetry breaking operators. We reproduce the result for
the Wilson coefficient of the
spin-orbit operator ${\cal O}_{SO}$ obtained in Ref.~\cite{BKP}
\begin{eqnarray}\label{cSO}
c_{SO}(\mu) = 2\left(\frac{\alpha_s(\mu)}{\alpha_s(m)}\right)^{-3/\beta_0} - 1\,.
\end{eqnarray}
This coefficient has been also obtained without an explicit calculation in  \cite{CKO}
from reparametrization invariance (RPI) arguments \cite{LM}. 
No such RPI constraints exist for the
coefficient of the Darwin term $c_D$, contrary to the argument of Ref.~\cite{CKO}. 
A discussion of the RPI constraints for the renormalized $1/m^2$ HQET Lagrangian is
given in Appendix B.

Finally, the coefficients of the remaining heavy-light four-quark operators (\ref{9}-\ref{11}) are
found to be
\begin{eqnarray}\label{c2hl}
c_2^{hl}(\mu) &=& -\frac{10}{21-4n_f}z^{-3/\beta_0} + \frac{5}{3-4n_f}z^{-6/\beta_0}
-\frac{15}{39-4n_f}\\
& & +\, \frac{20(99+96n_f-16n_f^2)}{(4n_f-39)(4n_f-21)(4n_f-3)}
z^{-(13-\frac43 n_f)/(2\beta_0)}\nonumber\\
c_3^{hl}(\mu) &=& 0\\\label{c4hl}
c_4^{hl}(\mu) &=& -\frac{4}{3(12-n_f)}z^{-3/\beta_0} + \frac{4}{3(15-2n_f)}z^{-6/\beta_0}
-\frac{4}{33-2n_f}\\
& & +\, \frac{4(-639+156n_f-8n_f^2)}{3(n_f-12)(2n_f-33)(2n_f-15)}
z^{-(22-\frac43 n_f)/(2\beta_0)}\,.\nonumber
\end{eqnarray}

A full one-loop determination of the coefficients $c^{hl}_i(\mu)$
requires an explicit matching calculation. This involves computing the box diagrams for
heavy-light quark scattering in QCD shown in Fig.1. When expanded in powers of $1/m$ the total result
for these diagrams is
\begin{eqnarray}\label{24}
I_{QCD} &=& \frac{ig^4}{(4\pi)^2}\left\{-\frac{3}{2\lambda^2}[\gamma_\mu t^a]_{\beta\alpha}
[\gamma^\mu t^a]_{\delta\gamma} + \frac{1}{m}\left[
-\frac{3\pi}{4\lambda}[\gamma_\mu t^a]_{\beta\alpha}[\gamma^\mu t^a]_{\delta\gamma} +
\frac{5\pi}{9\lambda}[\gamma_\mu\gamma_5t^a]_{\beta\alpha}[\gamma^\mu\gamma_5t^a]_{\delta\gamma} 
\nonumber\right.\right.\\
& &\hspace{-1cm}\left.\left.\,+
\frac{8\pi}{27\lambda}[\gamma_\mu\gamma_5]_{\beta\alpha}[\gamma^\mu\gamma_5]_{\delta\gamma}
\right]
+\frac{1}{m^2}\left[\frac94[\gamma_\mu t^a]_{\beta\alpha}[\gamma^\mu t^a]_{\delta\gamma} +
\frac{5}{12}\left(\ln\frac{\lambda^2}{m^2}-2\right)
[\gamma_\mu\gamma_5t^a]_{\beta\alpha}[\gamma^\mu\gamma_5t^a]_{\delta\gamma}
\right.\right.\nonumber\\
& &\hspace{-1cm}\left.\left.\,+
\frac29\left(\ln\frac{\lambda^2}{m^2}-2\right)
[\gamma_\mu\gamma_5]_{\beta\alpha}[\gamma^\mu\gamma_5]_{\delta\gamma}\right] + {\cal O}(1/m^3)
\right\}
\end{eqnarray}
To simplify the calculation we have taken a massless light quark scattering in the forward direction.
The infrared singularities have been regulated with a finite gluon mass $\lambda$.

\begin{figure}[b!]
\begin{tabular}{cc}  
{\includegraphics[height=5cm]{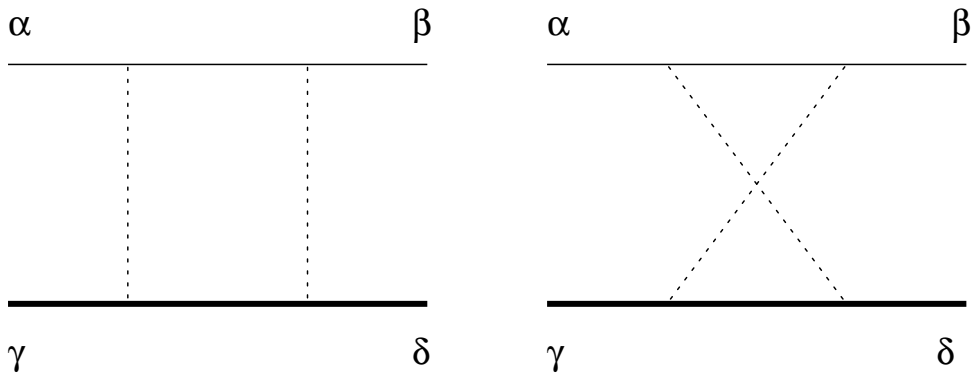}}
\end{tabular}
\caption{\label{fig1}
Diagrams contributing to heavy-light quark scattering which are needed for the matching
conditions for the four-quark operators $c_i^{hl}(m)$. The thick line denotes a heavy quark
and the dotted lines denote gluons.}
\end{figure}

A computation of the same diagrams using the effective theory Feynman rules for the heavy line
with insertions of all $1/m$ corrections up to second order gives, in the $\overline{MS}$ scheme,
\begin{eqnarray}\label{25}
I_{HQET} &=& \frac{ig^4}{(4\pi)^2}\left\{-\frac{3}{2\lambda^2}[\gamma_\mu t^a]_{\beta\alpha}
[\gamma^\mu t^a]_{\delta\gamma} + \frac{1}{m}\left[
-\frac{3\pi}{4\lambda}[\gamma_\mu t^a]_{\beta\alpha}[\gamma^\mu t^a]_{\delta\gamma} +
\frac{5\pi}{9\lambda}[\gamma_\mu\gamma_5t^a]_{\beta\alpha}[\gamma^\mu\gamma_5t^a]_{\delta\gamma} 
\nonumber\right.\right.\\
& &\hspace{-1cm}\left.\left.\,+
\frac{8\pi}{27\lambda}[\gamma_\mu\gamma_5]_{\beta\alpha}[\gamma^\mu\gamma_5]_{\delta\gamma}
\right]
+\frac{1}{m^2}\left[
\frac{5}{12}\left(\ln\frac{\lambda^2}{\mu^2}+2\right)
[\gamma_\mu\gamma_5t^a]_{\beta\alpha}[\gamma^\mu\gamma_5t^a]_{\delta\gamma}
\right.\right.\nonumber\\
& &\hspace{-1cm}\left.\left.\,+
\frac29\left(\ln\frac{\lambda^2}{\mu^2}+2\right)
[\gamma_\mu\gamma_5]_{\beta\alpha}[\gamma^\mu\gamma_5]_{\delta\gamma}\right] + {\cal O}(1/m^3)
\right\}
\end{eqnarray}

The first two terms in the $1/m$ expansion agree identically, as does the logarithmic dependence
on the infrared regulator $\lambda$ in the $1/m^2$ part. Imposing equality of these two expressions
requires adding the four-quark operators (\ref{8}-\ref{11}) to the effective theory with
coefficients
\begin{eqnarray}\label{26}
c_1^{hl}(\mu) &=& \frac{18g^2}{(4\pi)^2}\\
c_2^{hl}(\mu) &=& \frac{g^2}{(4\pi)^2}\left(\frac{10}{3}\log\frac{\mu^2}{m^2} -
\frac{40}{3}\right)\\
c_3^{hl}(\mu) &=& 0\\\label{29}
c_4^{hl}(\mu) &=& \frac{g^2}{(4\pi)^2}\left(\frac{16}{9}\log\frac{\mu^2}{m^2} -
\frac{64}{9}\right)\,.
\end{eqnarray}
The logarithmic terms agree, as they should, with the leading terms obtained by expanding the
LLA expressions Eqs.~(\ref{23b}), (\ref{c2hl})-(\ref{c4hl}).

One should point out that the constant terms in the full one-loop expressions for these
coefficients are scheme-dependent. The ${\cal O}(\alpha_s)$ terms become well-defined
only after renormalization-group improvement.
In this process the renormalization scheme dependence of the constant
terms in (\ref{26}-\ref{29}) is cancelled by the scheme dependence of a two-loop anomalous dimension
\cite{BFRS}.
However, there are reasons to expect the partial results (\ref{26}-\ref{29}) to provide a
good estimate of the full ${\cal O}(\alpha_s)$ correction, based on empirical evidence
\cite{BG} that the contribution of the two-loop anomalous dimension is often much smaller than the
constant term in the one-loop expression of the coefficient.

On the other hand, the logarithmic terms in (\ref{26}-\ref{29}) and the LLA sums
(\ref{23b},\ref{c2hl}-\ref{c4hl})
are scheme independent. However, the use of the leading-log resummed expression is appropriate
only when the logarithmic terms in (\ref{26}-\ref{29}) dominate clearly over the
constant terms. If this is not the case, then the use of the full one-loop results
(\ref{26}-\ref{29}),
with all the above limitations, is likely to be a better approximation to the true value of 
the Wilson coefficient. Such a situation is encountered in the case of the charm quark 
heavy mass expansion, as discussed in the next section.

\section{Application: heavy mesons hyperfine splitting}

The importance of considering the four-quark operators (\ref{8}-\ref{11}) becomes immediately
apparent if one notes that the equation of motion for the gluon field gives 
${\cal O}_D =  {\cal O}_1^{hl}$.
The coefficients of these two operators (\ref{23a}) and (\ref{23b}) get contributions of the
same order, such that neglecting ${\cal O}_1^{hl}$ would lead to an incomplete result.
Adding the contribution of  ${\cal O}_1^{hl}$  together with that of the Darwin term
gives, in the leading logarithmic
approximation, the ``effective'' coefficient Eq.~(\ref{cD}) for the local operator
of $O(1/m^2)$
\begin{equation}\label{34}
c_D(\mu) = \left(\frac{\alpha_s(\mu)}{\alpha_s(m)}\right)^{-6/\beta_0} -
\frac{308}{117}\left[\left(\frac{\alpha_s(\mu)}{\alpha_s(m)}\right)^{-13/(2\beta_0)} - 1
\right]\,.
\end{equation}
It is possible to eliminate ${\cal O}_1^{hl}$ from the effective Lagrangian
to all orders in $\alpha_s$, in favor of the Darwin operator, which contributes 
with coefficient $c_D(\mu) + c_1^{hl}(\mu) \to c_D(\mu)$ \cite{BM}.

More interesting from a practical point of view is the fact that the operators ${\cal O}_2^{hl}$ and
${\cal O}_4^{hl}$ break the heavy quark spin symmetry and thus contribute to the hyperfine
splittings of the heavy hadrons. The remainder of this Section is dedicated to a discussion of
these effects.

We first briefly review the derivation of the mass formula for the mass of a heavy hadron, including
the contributions of
the 4-quark operators (\ref{8})-(\ref{11}). The mass of a heavy hadron can be written as an 
expansion in
inverse powers of the heavy quark mass as \cite{Man,BSUV}
\begin{eqnarray}\label{mass}
m_{B^{(*)}} =m_b+\bar\Lambda  + \langle\mbox{T} {\cal H}_{int}\exp \left(-i\int\mbox{d}^4x
{\cal H}_{int}(x)\right)\rangle
\end{eqnarray}
with ${\cal H}_{int}=-{\cal L}_{int} = -\frac{1}{2m_b}{\cal L}_1 -\cdots$ the interaction
Hamiltonian.
The expectation value is
taken between eigenstates of the leading order HQET Lagrangian  which have the mass
$m_b+\bar\Lambda$.
This gives, including terms up to order $1/m_b^2$,
\begin{eqnarray}\label{ass}
m_{B^{(*)}} &=& m_b+\bar\Lambda - \frac{1}{2m_b}\langle {\cal L}_1\rangle - 
\frac{1}{(2m_b)^2}\left(\langle {\cal L}_2\rangle + 2(c_k(\mu))^2\langle {\cal O}_{kk}\rangle
\right.\\
&+&\left.\, c_k(\mu)c_m(\mu)\langle {\cal O}_{km}\rangle + 2(c_m(\mu))^2\langle {\cal O}_{mm}\rangle
+ \sum_{i=1}^4c_i^{hl}(\mu)\langle {\cal O}_i^{hl}\rangle\right) + \cdots\,.
\nonumber
\end{eqnarray}
The expectation values appearing in this formula can be parametrized as (using the
notations of Ref.~\cite{BSUV})
\begin{eqnarray}\label{para}
\langle {\cal L}_1\rangle &=& -c_k(\mu)\mu_\pi^2 - c_m(\mu)\frac{d_H}{3}\mu_G^2\\
\langle {\cal L}_2\rangle &=& c_D(\mu)\mu_D^3 + \frac{d_H}{3} c_{SO}(\mu)\mu_{SO}^3\\
\langle {\cal O}_{kk}\rangle &=& \mu_{\pi\pi}^3\\
\langle {\cal O}_{km}\rangle &=& \frac{d_H}{3}\mu_{\pi G}^3\\
\langle {\cal O}_{mm}\rangle &=& \mu_{GG}^3 + \frac12 \rho_S^3 + \frac16 d_H \rho_A^3\,,
\end{eqnarray}
with $d_H=3$ for a pseudoscalar meson and --1 for a vector meson.

The matrix elements of the four-quark operators (\ref{8})-(\ref{11}) can be estimated with the
help of the factorization approximation \cite{BSUV2}. (An alternative model-independent determination
of ${\cal O}_1^{hl}$ (based on a method proposed in \cite{C}) has been given in Ref.~\cite{CP}.)
This is done by first applying a Fierz transformation to bring the operator into the form
$(\bar h\Gamma' q)(\bar q\Gamma' h)$, after which the vacuum state is inserted between
the two currents.
Only color singlet currents will give a nonvanishing contribution. In this way we obtain
\begin{eqnarray}
& &\frac{1}{2m_B}\langle\bar B(v)|\frac{g^2}{2}\sum_q (\bar q\gamma_\mu\gamma_5 t^aq)(\bar h
\gamma^\mu\gamma_5 t^ah)|\bar B(v)\rangle = -\frac{2\pi\alpha_s}{3}f_B^2 m_B\\
& &\frac{1}{2m_B}\langle\bar B(v)|\frac{g^2}{2}\sum_q (\bar q\gamma_\mu\gamma_5q)(\bar h
\gamma^\mu\gamma_5 h)|\bar B(v)\rangle = -\frac{\pi\alpha_s}{2}f_B^2 m_B\\
& &\frac{1}{2m_B}\langle\bar B^*(v,\varepsilon)|\frac{g^2}{2}\sum_q (\bar q\gamma_\mu\gamma_5 t^aq)
(\bar h\gamma^\mu\gamma_5 t^ah)|\bar B^*(v,\varepsilon)\rangle = \frac{2\pi\alpha_s}{9}f_B^2 m_B\\
& &\frac{1}{2m_B}\langle\bar B^*(v,\varepsilon)|\frac{g^2}{2}\sum_q (\bar q\gamma_\mu\gamma_5q)
(\bar h\gamma^\mu\gamma_5 h)|\bar B^*(v,\varepsilon)\rangle = \frac{\pi\alpha_s}{6}f_B^2 m_B\,.
\end{eqnarray}

We used here the following relations which are valid in the heavy mass limit
\begin{eqnarray}
& &\langle\bar B(v)|\bar h\gamma_\mu\gamma_5q|0\rangle = if_Bm_B v_\mu\\
& &\langle\bar B^*(v,\varepsilon)|\bar h\gamma_\mu q|0\rangle = f_{B}m_{B} \varepsilon_\mu^*\,.
\end{eqnarray}

Combining these relations one obtains the following expression for the hyperfine splitting
\begin{eqnarray}
m_{B^*} - m_B &=& -\frac{1}{2m_b}c_m(\mu)\frac{4\mu_G^2}{3} + \frac{4}{3(2m_b)^2}\left(
c_{SO}(\mu)\mu_{SO}^3 + c_k(\mu)c_m(\mu)\mu_{\pi G}^3  \right.\\
& &\left.\, + (c_m(\mu))^2 \rho_A^3 - c_2^{hl}(\mu)\frac{2\pi\alpha_s}{3}
f_B^2 m_B -  c_4^{hl}(\mu)\frac{\pi\alpha_s}{2}f_B^2 m_B\right) + {\cal O}(1/m_b^3)\,.\nonumber
\end{eqnarray}

\begin{center}
\begin{tabular}{|c|c|c|c|c|}
\hline
& \multicolumn{2}{|c|}{LLA}
& \multicolumn{2}{c|}{One-loop}\\
\cline{2-5}
 & $\mu=0.5$ GeV
 & $\mu=1$ GeV
 & $\mu=0.5$ GeV
 & $\mu=1$ GeV\\[0.1cm]
\hline
\hline
& & & &\\[-0.3cm]
$c_2^{hl}(\mu/m_c)$ & $-0.268$ & $-0.074$ & $-0.653$ & $-0.503$ \\[0.4cm]
\hline
& & & &\\[-0.3cm]
$c_4^{hl}(\mu/m_c)$ & $-0.117$ & $-0.038$ & $-0.348$ & $-0.268$ \\[0.4cm]
\hline
& & & &\\[-0.3cm]
$c_2^{hl}(\mu/m_b)$ & $-0.401$ & $-0.232$ & $-0.577$ & $-0.483$ \\[0.4cm]
\hline
& & & &\\[-0.3cm]
$c_4^{hl}(\mu/m_b)$ & $-0.156$ & $-0.105$ & $-0.308$ & $-0.268$ \\[0.4cm]
\hline
\end{tabular}
\end{center}
\begin{quote} {\bf Table 1.}
The coefficients of the spin-symmetry violating four-quark operators for the 
charm and bottom cases,
at two values of the factorization scale $\mu=0.5$ GeV and $\mu=1$ GeV.
\end{quote}

This relation can be written only in terms of observable quantities by expressing the quark mass
in terms of the pseudoscalar meson mass and multiplying with $m_B + m_{B^*}$
\begin{eqnarray}\label{hyper}
m_{B^*}^2 - m_B^2 &=& - c_m(\mu)\frac{4\mu_G^2}{3} + \frac{4}{3(2m_B)}\left(
-2\bar\Lambda c_m(\mu) \mu_G^2 + c_{SO}(\mu)\mu_{SO}^3 + c_k(\mu)c_m(\mu)\mu_{\pi G}^3\right.\\
& &\left.\,+ (c_m(\mu))^2 \rho_A^3 - c_2^{hl}(\mu)\frac{2\pi\alpha_s}{3}
f_B^2 m_B -  c_4^{hl}(\mu)\frac{\pi\alpha_s}{2}f_B^2 m_B\right) + {\cal O}(1/m_B^2)\,.\nonumber
\end{eqnarray}

In Table 1 we present values for the coefficients of the two four-quark operators which contribute
to the hyperfine splittings. They are given in both the leading-log approximation and the
``full'' one-loop form at two different values of the factorization scale $\mu=0.5$ and 1 GeV.
In computing these values we used $\Lambda_{QCD}=250$ MeV and $m_c=1.39$ GeV, $m_b=4.8$ GeV. For the
bottom hadron case we neglected the change in the running of $\alpha_s$ across the charm 
threshold, which
gives a negligible error.
The one-loop result suggests that the leading-log approximation is very poor for the
charm case, where the logarithmic term accounts for at most 30\% of the total correction.
The situation is slightly better in the bottom quarks, where its contribution is enhanced 
to about 50\%.

For the purpose of illustration we show in Table 2 the combined contributions of
these two operators
to the hyperfine splitting of the D and B mesons using for their coefficients both approximation
methods.
For the reasons discussed above we tend to
favor the full one-loop results over those obtained from the leading log approximation.
In computing the matrix elements of the operators in Table 2 we have used the
hadronic parameters $f_D=170$ MeV, $f_B=180$ MeV, $m_D=1.97$ GeV, $m_B=5.28$ GeV.

\begin{center}
\begin{tabular}{|c|c|c|c|c|c|c|}
\hline
& & & & & &\\[-0.3cm]
$\mu_{fact}$ & $\langle {\cal O}_2^{hl}\rangle_{D^*-D}$
& $\langle {\cal O}_4^{hl}\rangle_{D^*-D}$
& $\langle {\cal O}_2^{hl}\rangle_{B^*-B}$
& $\langle {\cal O}_4^{hl}\rangle_{B^*-B}$
& $\Delta_D$ (MeV) & $\Delta_B$ (MeV)\\[0.3cm]
\hline
\hline
& & & & & & \\[-0.3cm]
0.5 GeV & (543 MeV)$^3$ & (493 MeV)$^3$ & (783 MeV)$^3$ & (712 MeV)$^3$ 
& 7.4(LLA)& 2.7(LLA)\\
& & & & & 18.9(1-l.) & 4.2(1-l.)\\[0.3cm]
\hline
& & & & & & \\[-0.3cm]
1.0 GeV & (431 MeV)$^3$ & (392 MeV)$^3$ & (622 MeV)$^3$ & (565 MeV)$^3$ 
& 1.1(LLA) & 0.8(LLA)\\
& & & & & 7.3(1-l.) & 1.8(1-l.)\\[0.3cm]
\hline
\end{tabular}
\end{center}
\begin{quote} {\bf Table 2.}
Matrix elements of the four-quark operators ${\cal O}_2^{hl}$ and ${\cal O}_4^{hl}$ in the
factorization approximation for two different values of the factorization scale $\mu=0.5$ and
1 GeV. In the last two columns the total contribution of these two operators to the
hyperfine splittings of the $D$ and $B$ mesons $\Delta_i = m_{H_i^*} - m_{H_i}$
is shown in both the leading-log approximation and using
the one-loop result (1-l.) for their coefficients.
\end{quote}

The relation (\ref{hyper}) contains the as yet unknown parameters $\mu_{SO}^3, \mu_{\pi G}^3$
and $\rho_A^3$.
The first parameter vanishes exactly for $S$-wave mesons in potential models as the corresponding
operator
(\ref{5}) corresponds to the spin-orbit interaction energy of the heavy quark \cite{BSUV}. The 
remaining two parameters are not easy to estimate in a model-independent way. The contributions 
of the four-quark
operators are shown in the last two columns of Table 2. Their total contribution is positive and
amounts to about 10-20 MeV in the charm case and 1-5 MeV in the bottom case. 

Comparing with the measured hyperfine splittings one can see that a positive $1/m_B$ correction to
(\ref{hyper}) agrees with the sign of the correction to the heavy mass scaling law.
Unfortunately, the lack of information on the precise values of $\mu_{SO}^3,
\mu_{\pi G}^3, \rho_A^3$ prevents us from making a more quantitative analysis of the 
deviations from leading order scaling. 

Finally we note the relevance of these corrections for the model-independent determination of the
matrix element $\mu_G^2$ from the measured hyperfine splittings in the B system. From (\ref{hyper})
we obtain 0.358 GeV$^2$ = $\mu_G^2 + 0.026 + 0.021$ + terms proportional to $\mu_{SO}^3/m_B$ and
$\mu_{\pi G}^3/m_B$. The two numbers on the right-hand side correspond to $\bar\Lambda\mu_G^2$ and
respectively to the four-quark operators' contribution. In this estimate we used $\bar\Lambda=400$
MeV and $\mu_G^2$ =0.35 GeV$^2$. This shows that the theoretical error of this determination
coming from the $1/m_B$ terms in (\ref{hyper}) may be as high as 13\%. 

\begin{acknowledgements}
D.~P. is grateful to A.~Czarnecki, C.~D.~L\"u and D.~X.~Zhang for discussions and 
acknowledges support of the Deutsche Forschungsgemeinschaft (DFG) while part of this work 
was done. He thanks A.~Manohar for discussions which helped identify errors in a previous
version of this paper.
J.~G.~K. acknowledges partial support from BMFT, FRG under contract 06MZ566. J.~C.~R. acknowledges
the support of the Alexander von Humboldt Foundation.
\end{acknowledgements}

\newpage
\appendix
\section*{Appendix A}

A complete study of the renormalized HQET Lagrangian at order $1/m^2$ requires also the
computation of the coefficients of operators which vanish by the equation of motion of the
heavy quark field. There are altogether six such operators: three local operators which will be
chosen as in \cite{BKP}
\begin{eqnarray}\label{B1}
{\cal O}_8 &=& \frac{ig}{2}\bar h\sigma_{\mu\nu}F^{\mu\nu}(v\cdot D)h +
\frac{ig}{2}\bar h(v\cdot D)\sigma_{\mu\nu}F^{\mu\nu}h\\
{\cal O}_9 &=& i\bar hD^2(v\cdot D)h + i\bar h(v\cdot D)D^2h\\\label{B3}
{\cal O}_{10} &=& \bar h(iv\cdot D)^3h\,,
\end{eqnarray}
and three nonlocal operators consisting of time-ordered products of dimension 5 operators
\begin{eqnarray}\label{B4}
{\cal O}_{ee} &=& \frac{i}{2}\int\mbox{d}^4x\,\mbox{T}[\bar h(iv\cdot D)^2h](x)
[\bar h(iv\cdot D)^2h](0)\\
{\cal O}_{ke} &=& i\int\mbox{d}^4x\,\mbox{T}[\bar h(iD)^2h](x)
[\bar h(iv\cdot D)^2h](0)\\
{\cal O}_{me} &=& i\int\mbox{d}^4x\,\mbox{T}[\bar h\frac{g}{2}\sigma\cdot Fh](x)
[\bar h(iv\cdot D)^2h](0)\,.
\end{eqnarray}
They will be combined together in a vector ${\cal H}_e$ defined as
\begin{eqnarray}\label{B7}
{\cal H}_e = \mbox{column}({\cal O}_8\,, {\cal O}_9\,, {\cal O}_{10}\,,
{\cal O}_{ee}\,, {\cal O}_{ke}\,, {\cal O}_{me})\,.
\end{eqnarray}

These operators mix only with the operators in ${\cal H}$ (\ref{19}). The renormalized operators
${\cal H}$ and ${\cal H}_e$ satisfy a renormalization group equation which can be written in
matrix form as
\begin{eqnarray}\label{B8}
\mu\frac{\mbox{d}}{\mbox{d}\mu} 
\left( \begin{array}{c}
{\cal H}\\
{\cal H}_e\end{array} \right) +
\left( \begin{array}{cc}
A & G\\
0 & H\end{array} \right)
\left( \begin{array}{c}
{\cal H}\\
{\cal H}_e\end{array} \right) = 0\,.
\end{eqnarray}
The block $A$ has been given in (\ref{22a}), and the remaining ones are given by
\begin{eqnarray}
G = \frac{g^2}{(4\pi)^2}
\left( \begin{array}{cccccc}
0 & 0 & -\frac{16}{3} & 0 & 0 & 0 \\
3 & 0 & 0 & 0 & 0 & 0 \\
0 & \frac{32}{3} & \frac{128}{3} & 0 & -\frac{32}{3} & 0 \\
-\frac{14}{3} & 0 & 0 & 0 & 0 & -\frac{32}{3} \\
9 & 0 & -\frac{32}{3} & 0 & 0 & 0 \end{array} \right)\,,\\
H = \frac{g^2}{(4\pi)^2}
\left( \begin{array}{cccccc}
12 & 0 & 0 & 0 & 0 & 0 \\
0 & 0 & -\frac{32}{3} & 0 & 0 & 0 \\
0 & 0 & \frac{16}{3} & 0 & 0 & 0 \\
0 & 0 & \frac{16}{3} & \frac{32}{3} & 0 & 0 \\
0 & -\frac{16}{3} & -32 & -\frac{64}{3} & \frac{16}{3} & 0 \\
-\frac23 & 0 & 0 & 0 & 0 & \frac{34}{3}\end{array} \right)\,.
\end{eqnarray}

At the scale $\mu=m$ the coefficients of the operators in (\ref{B7}) are given by
\begin{equation}
c_e = \mbox{column}(0\,,0\,,0\,,1\,,-1\,,1)\,.
\end{equation}
Integrating the RG equation (\ref{B8}) with this initial condition one obtains the following
expressions for the coefficients of the three local operators in (\ref{B7}) in the leading log
approximation
\begin{eqnarray}\label{B12}
c_8(\mu) &=& -2z^{-3/\beta_0} + 3z^{-17/(3\beta_0)} - \frac54 z^{-6/\beta_0} + \frac14
- \frac{9}{2\beta_0}z^{-6/\beta_0}\ln z\\\label{B13}
c_9(\mu) &=& 3z^{-8/(3\beta_0)} - 3\\
c_{10}(\mu) &=& -\frac35 z^{-6/\beta_0} - \frac{65}{9} z^{-2/\beta_0} + 9 z^{-16/(3\beta_0)}
+ \frac{23}{30} z^{-8/(3\beta_0)} - \frac{35}{18}\,.\label{B14}
\end{eqnarray}
We denoted here $z=(\alpha_s(\mu)/\alpha_s(m))$. In deriving (\ref{B14}) we used the result
derived in Sec.~III for the Wilson coefficient of the Darwin term.
It is important to emphasize that these coefficients
are (just as $c_e(\mu)$ in (\ref{4})) gauge dependent. Their expressions given above correspond to
the Feynman gauge.

\newpage
\section*{Appendix B}

The matching relation connecting the heavy quark field $Q$ in QCD with the effective theory field
$h$ is \cite{KT,BKP}
\begin{eqnarray}\label{A1}
Q(x) = \Omega\exp(-imv\cdot x\vsl) h(x)\,,
\end{eqnarray}
with
\begin{eqnarray}\label{A2}
\Omega = \exp\left(\frac{i\Dslp}{2m}\right)\exp\left(\frac{1}{4m^2}\left[
(i\Dsl_\parallel)(i\Dslp) + (i\Dslp)(i\Dsl_\parallel)\right]\right)\cdots
\end{eqnarray}
and $\Dsl_\parallel = v\cdot D\vsl$, $\Dslp=\Dsl-\Dsl_\parallel$. The field $h$ contains
both ``upper'' and ``lower'' components $h_\pm$, satisfying $\vsl h_\pm = \pm h_\pm$. One has
\begin{eqnarray}
\exp(-imv\cdot x\vsl)h = e^{-imv\cdot x}\frac{1+\vsl}{2}h + e^{imv\cdot x}\frac{1-\vsl}{2}h
= e^{-imv\cdot x}h_+ + e^{imv\cdot x}h_-\,.
\end{eqnarray}
It can be shown \cite{KT} that the field transformation (\ref{A2}) decouples the two
components $h_+$ and $h_-$. The HQET Lagrangian (\ref{1}-\ref{3}) used in the main text refers
only to the $h_+$ part.

The most general form for the $1/m^2$ term in the HQET Lagrangian includes, in addition to the
operators introduced so far, also the operators
\begin{eqnarray}\label{A4}
{\cal O}_8 &=& \frac{ig}{2}\bar h\sigma_{\mu\nu}F^{\mu\nu}(v\cdot D)h +
\frac{ig}{2}\bar h(v\cdot D)\sigma_{\mu\nu}F^{\mu\nu}h\\
{\cal O}_9 &=& i\bar hD^2(v\cdot D)h + i\bar h(v\cdot D)D^2h\\\label{A6}
{\cal O}_{10} &=& \bar h(iv\cdot D)^3h\,.
\end{eqnarray}
Their renormalization properties are studied in the Appendix A.

The purpose of this Appendix is to show that requiring the invariance of the HQET Lagrangian 
(\ref{1}) under a small change in the velocity $v$ (the so-called reparametrization invariance
\cite{LM}) fixes the coefficients $c_{SO}(\mu)$ \cite{CKO} and $c_9(\mu)$. However, no constraint
for $c_D(\mu)$ is obtained in this way, in contrast to \cite{CKO}.

We start by computing the change in the $h_+$ field under an infinitesimal change of the velocity
parameter $v\to v'=v+\Delta v$. This can be obtained from (\ref{A1}) by taking into account the
invariance of the QCD field under this transformation $\delta Q=0$ as
\begin{eqnarray}\label{A7}
\delta h_+ = \delta\left[ \frac{1+\vsl}{2}\exp(imv\cdot x)\Omega^{-1}\right]\Omega
\exp(-imv\cdot x\vsl)h\,.
\end{eqnarray}
Explicitly, to first order in $\Delta v$ and up to second order in $1/m$,
\begin{eqnarray}\label{A8}
\delta h_+ = \Delta v^\mu\left\{ imx_\mu + \frac12 \gamma_\mu + \frac{i}{4m}
\left[\gamma_\mu \Dslp + D_\mu\right] + {\cal O}(1/m^2)\right\} h_+\,.
\end{eqnarray}
Note that no negative component field $h_-$ is introduced by this transformation.

The variation of the HQET Lagrangian (\ref{1}-\ref{3}) is obtained by the variation of the
effective field (\ref{A8}) and the variation of the $v$-dependent operators, including the
operators (\ref{A4}-\ref{A6}). After some algebra one obtains
\begin{eqnarray}\label{A9}
\delta {\cal L} &=& (1-c_k(\mu)) \bar h_+i\Delta v\cdot Dh_+
-\frac{1}{m}\left\{-(c_e+c_9-1)\bar h_+(\Delta v\cdot D)(v\cdot D)h_+\right.\\
&-&\left.\,\frac14\left(c_e+1+c_9+c_m+\frac{c_{SO}+1}{2}\right)
\bar h_+ig\gamma\cdot \Delta v\gamma^\nu v^\mu F_{\mu\nu} h_+\right.\nonumber\\
&-&\left.\,\frac14\left(c_e+1+c_9-c_m-\frac{c_{SO}+1}{2}\right)
\bar h_+ig\gamma^\nu\gamma\cdot \Delta v v^\mu F_{\mu\nu} h_+{\cal O}(1/m^2)\right\}\nonumber\,.
\end{eqnarray}
This will vanish for any $\Delta v$ provided that the following identities hold
\begin{eqnarray}\label{A10}
c_k(\mu) &=& 1\\\label{A11}
c_9(\mu) &=& -1 - c_e(\mu)\\\label{A12}
c_{SO}(\mu) &=& -1 -2 c_m(\mu)\,.
\end{eqnarray}

The first constraint has been given in \cite{LM} and is perhaps the best known application of
the reparametrization invariance principle. The constraint (\ref{A12}) has been presented in
\cite{CKO} and its prediction for $c_{SO}$ agrees with the explicit calculation in \cite{BKP}
(see (\ref{cSO})). The relation (\ref{A11}) together with (\ref{6}) predicts the following value
for $c_9(\mu)$
\begin{eqnarray}
c_9(\mu) = 3\left[\left(\frac{\alpha_s(\mu)}{\alpha_s(m)}\right)^{-8/(3\beta_0)} - 1\right]\,.
\end{eqnarray}
This agrees with the expression of this coefficient obtained by direct computation in leading
log approximation (\ref{B13}).

\end{document}